\documentclass[twocolumn,preprintnumbers,showkeys,amsmath,amssymb,prb]{revtex4}

\usepackage{graphicx}
\usepackage{dcolumn}
\usepackage{bm}

\newcommand{\ro}{\bm{\rho}}

\newcommand{\bn}{\bm{\nabla}}
\begin{document}

\preprint{To be reported at 17$^{\rm th}$ Int. Conference on High Magnetic Fields,
W\"{u}rzburg, Germany, July 30 - August 4, 2006}

\title{Many-body Effects in Landau Levels: Non-commutative Geometry \\ 
and Squeezed Correlated States}

\author{Alexander B. Dzyubenko}\email{adzyubenko@csub.edu}
\homepage{http://www.csub.edu/~adzyubenko}

\affiliation{
Department of Physics, California State University at Bakersfield, Bakersfield, CA 93311, USA}
\affiliation{General Physics Institute, Russian Academy of Sciences, Vavilova 38, Moscow 119991, Russia}

\date{\today}

\begin{abstract}
We discuss symmetry-driven squeezing and coherent states of few-particle systems 
in magnetic fields. An operator approach using canonical transformations 
and the $SU(1,1)$ algebras is developed for considering Coulomb correlations 
in the lowest  Landau levels.
\end{abstract}


\maketitle

\section{Introduction}

Coulomb interactions in strong magnetic fields are relevant in the various physical 
contexts such as, for example, the behavior of ions in atmospheres of neutron stars, 
electrons and holes in semiconductors, and charged quasiparticles 
in the Fractional Quantum Hall Effect regime. One of the general aspects of the problem 
are considerations of the relevant symmetries. 
In this work, we discuss an operator method that allows one to maintain both
axial and magnetic translations geometric symmetries 
in two-dimensional (2D) systems in Landau Levels (LL). We also  
establish a connection with the dynamical $SU(1,1)$ symmetry.

\section{Charged $e$-$h$ Systems}
            \label{Sec:Charged}

Let us consider a 2D system of two oppositely 
charged particles $-q_1 <0$ and $q_2 >0$, which
we will denote as an ``electron'' and a ``hole'', respectively.  
The total charge is negative,  
$-Q = q_2 - q_1 < 0$.
The operator of magnetic translations (MT) is of the form
$\hat{\bf K} = -i\hbar\bn_{1} + -i\hbar\bn_{2} - {\bf B} 
\times (q_1 {\bf r}_1 - q_2 {\bf r}_2 )/2c$,
where the symmetric gauge is used.\cite{Avron78,Dzyubenko01}
The MT group  is non-commutative, and the dimensionless MT operator
$\hat{\bf k} =\hat{\bf K} L_B/\hbar$
has canonically conjugate components,
$ [ \hat{k}_x , \hat{k}_y ] = i$, 
where $L_B \equiv \sqrt{ \hbar c / QB }$ is the effective magnetic length.
This allows one to introduce a pair of Bose ladder operators
for the whole 
system,\footnote{For a multiparticle $e$-$h$ system, 
charges $-q_1$, $q_2$ and coordinates ${\bf r}_1$, ${\bf r}_2$
correspond to the total charges and center-of-charge coordinates
of the $e$- and $h$- subsystems.\protect\cite{Dzyubenko01} }
$\tilde{B}_e^{\dag} = - i\hat{k}_{-}/\sqrt{2}$ and 
$\tilde{B}_e =  i\hat{k}_{+}/\sqrt{2}$ such that 
$ [ \tilde{B}_e , \tilde{B}_e^{\dag} ] = 1$,
here $\hat{k}_{\pm} = \hat{k}_x \pm i \hat{k}_y$.
In terms of intra-LL operators of individual particles, we have\cite{Dzyubenko01}
\begin{eqnarray}
        \label{lad1}
   \tilde{B}_e^{\dag} & = & - \frac{i\hat{k}_{-}}{\sqrt{2}}
        = u B_e^{\dag} - v B_h \quad , \\
          u & = & \sqrt{\frac{q_1}{Q}}  \quad , \quad 
          v = \sqrt{\frac{q_2}{Q}} \quad .
\end{eqnarray}
The second linearly independent annihilation operator is 
$\tilde{B}_h  = u B_{h} - v B_{e}^{\dag}$ 
so that we have two pairs of Bose ladder operators: 
$[\tilde{ B}_h, \tilde{B}_h^{\dag}]=[\tilde{ B}_e, \tilde{B}_e^{\dag}]=1$, 
$[\tilde{B}_h, \tilde{B}_e^{\dag}]=0$, and  
$[\tilde{B}_h, \tilde{B}_e]=0$.
This is in fact Bogoliubov canonical transformation 
\begin{eqnarray}
        \label{B-Brho}
\left( \begin{array}{c}
                  \tilde{B}^{\dag}_{e} \\
                   \tilde{B}_{h}
                           \end{array} \right) & = &
 \left( \begin{array}{c}
                   \tilde{S} B^{\dag}_{e} \tilde{S}^{\dag} \\
                   \tilde{S} B_{h} \tilde{S}^{\dag}
                           \end{array} \right) 
   =  \hat{U} \left( \begin{array}{c} B^{\dag}_{e} \\
                                   B_{h}
                 \end{array} \right)  \quad , \\
 \hat{U} & \equiv &     \left( \begin{array}{rr}
            \cosh\Theta & -\sinh\Theta  \\
           -\sinh\Theta &  \cosh\Theta
                       \end{array} \right)  \quad
\end{eqnarray}
performed by the unitary operator $\tilde{S} =   \exp ( \Theta \tilde{\cal L} )$
with the generator $\tilde{\cal L} =   B^{\dag}_{h} B^{\dag}_{e} - B_e B_h$;
here $\Theta$ is the transformation parameter with
$u= \cosh\Theta$, $v= \sinh\Theta$. 
This transformation introduces new quasiparticles with coordinates 
\begin{equation}
        \label{R1}
 {\bf R}_1 = \frac{ q_1 {\bf r}_1 - q_2 {\bf r}_2 }{Q} 
\quad , \quad 
 {\bf R}_2  = \frac{\sqrt{q_1 q_2}}{Q} \, ( {\bf r}_2 - {\bf r}_1 )   \quad ,
\end{equation}
in which the transformed operators assume the standard forms\cite{Dzyubenko01}
\begin{eqnarray}
   \tilde{B}^{\dag}_{e} &=& \frac{1}{\sqrt{2}} \left( \frac{Z^{\ast}_1 }{2L_B} -
                      2L_B \frac{\partial}{\partial Z_1} \right)      \quad , \\
\tilde{B}^{\dag}_{h} &=& \frac{1}{\sqrt{2}} \left( \frac{Z_2 }{2L_B} -
                      2L_B \frac{\partial}{\partial Z_2^{\ast}} \right)   \quad , 
\end{eqnarray}
where the 2D complex variables $Z_i = X_i + i Y_i$ are used.
Note that the interaction potential 
$U_{\rm int}=U( {\bf r}_1 - {\bf r}_2)$ 
does not depend on ${\bf R}_1$. 
The MT operator is diagonal in the new representation,
${\bf k}^2 = 2\tilde{B}^{\dag}_{e} \tilde{B}_{e}+1$.
It has the discrete spectrum $2k+1$, 
where the oscillator quantum numbers $k=0, 1, 2, \ldots$
determine the position of a guiding center of a charged system 
in ${\bf B}$.\cite{Avron78,Dzyubenko01}
A complete basis of states in zero LL compatible with both axial and translational
symmetries is given by 
\begin{equation}
            \label{basis}
       \frac{ \tilde{B}_e^{\dag}\mbox{}^k
              \tilde{B}_h^{\dag}\mbox{}^m}
                {\sqrt{k!m!}} |\tilde{0} \rangle  
  \equiv  |\widetilde{k m}\rangle \quad ,
\end{equation}
where $| \tilde{0} \rangle = \tilde{S} | 0 \rangle $
is the transformed vacuum and state 
$|\widetilde{ k m} \rangle$  has total angular
momentum projection $M_z=m-k$. 
The energy spectrum is degenerate with respect to $k$.
Therefore, it is sufficient to consider only the states with $k=0$
from (\ref{basis});  we denote such states as $|\tilde{ m}\rangle$.
The above procedure removes one degree of freedom and corresponds to a 
possible partial separation of variables in magnetic fields. 

For, e.g., the Coulomb interaction
$U_{\rm int}=- q_1 q_2 /|{\bf r}_1 - {\bf r}_2|$, 
the eigenergies in the lowest LL can be calculated
analytically as expectation values: 
\begin{eqnarray}
          \label{Um}
   U_m &=&  \langle \tilde{m}| U_{\rm int} | \tilde{m} \rangle  \\
     \nonumber
       & = & - E_{0} \left(\frac{q_2}{q_1}\right)^{m+\frac{3}{2}}
              \sum_{k=0}^{m} C_m^k
      \frac{\Gamma(k+\frac{1}{2})}{\sqrt{\pi} \, k!}
       \left(\frac{q_1-q_2}{q_2}\right)^{k}  \quad ,
\end{eqnarray}
where $E_{0} = \sqrt{\frac{\pi}{2}} \frac{q_1^2}{l_{B1}}$,
$l_{Bi}=(\hbar c/q_iB)^{1/2}$ are magnetic lengths.
Eigenenergies (\ref{Um}) are shown in Fig.~1 
for several values of parameter $\epsilon = (q_2/q_1)^{1/2} < 1 $. 
The spectra are completely discrete. However, in the limit $q_2 \rightarrow q_1$ 
($\epsilon \rightarrow 1^{-0}$) the spectra become {\em quasicontinuous}
and fill in the 2D neutral magnetoexciton band of width $E_0$.\cite{Lerner80}

\begin{figure}[t]
\includegraphics[width=.48\textwidth]{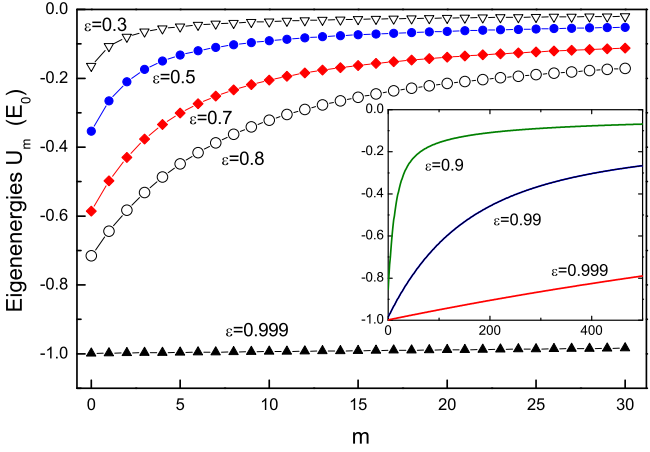}
\caption{
Eigenergies (\protect\ref{Um}) of two interacting charges $-q_1 <0$ and $q_2>0$
in zero LL for several values of parameter $\epsilon = (q_2/q_1)^{1/2}$.  
}
\end{figure}

In the terminology of quantum optics, the transformed vacuum 
$|\tilde{0}\rangle = \tilde{S}|0\rangle$
is a two-mode squeezed state.\cite{Scully97} 
For particles in a magnetic field squeezing
has a direct geometrical meaning.\cite{Dzyubenko01} 
Indeed, in the coordinate representation we have  
\begin{equation}
        \label{Psim1e}
   \langle {\bf r}_1,{\bf r}_2 | \tilde{0} \rangle =
   \frac{\sqrt{1-\epsilon^2}}{2\pi l_{B1} l_{B2} }
  \exp\left(
   - \frac{{\bf r}_1^2}{4l_{B1}^2}
   - \frac{{\bf r}_2^2}{4l_{B2}^2}
   + \frac{ \epsilon z_1^*z_2}{2l_{B1}l_{B2}} \right)  \quad .
\end{equation}
Using Eq.~(\ref{Psim1e}), the  probability distribution can be presented 
in the following form
\begin{eqnarray}
        \label{Psim1-prob}
  \left| \langle {\bf r}_1,{\bf r}_2 | \tilde{0} \rangle  \right|^2 
                      & \sim &  \frac{1}{\sigma_+ \sigma_-} 
  \exp\left( - \frac{\ro_+^2}{\sigma_+^2} - \frac{\ro_-^2}{ \sigma_-^2} \right)
   \quad , \\
\ro_{\pm} & = & \frac{{\bf r}_1}{l_{B1}} \pm \frac{{\bf r}_2}{l_{B2}}  \quad ,
\end{eqnarray}
where
$\sigma_{\pm}^2 = 4/(1 \mp \epsilon)$. 
This shows that in the new vacuum state $| \tilde{0}\rangle$ the distribution probability
for the difference coordinate $\ro_-$ is squeezed at the expense 
of the sum coordinate $\ro_+$, see Fig.~2. 
\begin{figure}[b]
\includegraphics[width=.48\textwidth]{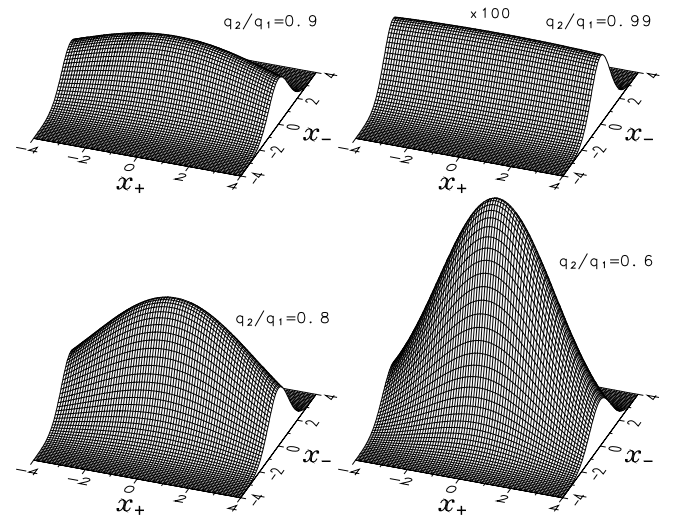}
\caption{
Probability distribution (\protect\ref{Psim1-prob})
for the $x$-components of coordinates 
$\protect\ro_+$ and  $\protect\ro_-$ (at $y_+=y_-=0$) 
for several different values of the charge ratio
$q_2/q_1$.
}
\end{figure}
%
for the $x$-components of coordinates 
$\protect\ro_+$ and  $\protect\ro_-$ (at $y_+=y_-=0$) 
for several different values of the charge ratio
$q_2/q_1$.
Note that in the limit
$q_2 \rightarrow q_1$ ($\epsilon \rightarrow 1^{-0}$)
the relative coordinate distribution becomes maximally
squeezed while
the center-of-charge coordinate becomes {\em extended},
$\sigma_+ \rightarrow \infty$.
This is because the components of 
$\ro_1$ and $\ro_2$ do not commute in the lowest LL approximation, 
see Sec.~\ref{Sec:Neutral} below.

\section{Neutral $e$-$h$ Systems}
                     \label{Sec:Neutral} 

For a neutral system $q_2 = - q_1=q$ the MT operator is given by 
$\hat{\bf K} = -i\hbar\bn_{1} + -i\hbar\bn_{2} - q{\bf B} \times ({\bf r}_1 -{\bf r}_2 )/2c$.
Its components commute $[ \hat{K}_x,\hat{K}_y ] = 0$ 
so that the MT group is abelian. Therefore, the states of a neutral magnetoexciton (MX)
can be labeled by the magnetic momentum 
${\bf K}=(K_x,K_y)$.\cite{Lerner80,Laughlin84}
The ground state in zero LL is a  ${\bf K}=0$ state, 
which can be presented as a squeezed two-mode vacuum
\begin{equation}
        \label{K0}
  | {\bf K} = 0 \rangle =  S | 0 \rangle  
          \quad , \quad  
        S =  \exp \left( B_e^{\dag} B_h^{\dag} \right) \quad .
\end{equation}
The coordinate representation is given by\cite{Lerner80}
\begin{equation}
        \label{coorK01}
 \langle {\bf r}_1 {\bf r}_2| {\bf K} = 0 \rangle =
   \exp \left(
     - \frac{ {\bf r}_1^2 + {\bf r}_2^2 - 2z_1^* z_2}{4l_B^2} \right) \quad .
\end{equation}
This is a coherent state of an infinite number of electron
and hole states in zero LL.\cite{Laughlin84} 
A state with a finite momentum ${\bf K}=(K_x,K_y)$ is given by
\begin{eqnarray}
        \label{K}
  | {\bf K} \rangle &=&  | K_x , K_y \rangle
                   = S({\bf K})| 0 \rangle  \quad , \\
        \label{2sqzK}
        S({\bf K})  &=&
   \exp\left[ \left( B_e^{\dag} + \frac{ik_-}{\sqrt{2}} \right)
              \left( B_h^{\dag} + \frac{ik_+}{\sqrt{2}} \right) \right]  \quad 
\end{eqnarray}
and is a two-mode squeezed displaced vacuum 
state.\footnote{Note that for, e.g., 
the electron operator, 
$D_e(\alpha) B_e^{\dag} D_e^{\dag}(\alpha)=B_e^{\dag} + \alpha$,
where a well known displacement operator\cite{Scully97,Perelomov86}                  
is given by 
$D_e(\alpha)=\exp(\alpha B_e^{\dag} - \alpha^* B_e)$. }
Expectation value of the relative coordinate $ {\bf r} = {\bf r}_1 - {\bf r}_2$
in state (\ref{K}) is given by 
$\langle {\bf K} |{\bf r} |{\bf K} \rangle = \frac{{\bf B}}{B} \times 
{\bf K} \, \hbar^{-1}l_B^2$.\cite{Lerner80}
Also, the zero-momentum state (\ref{coorK01}) can be considered to be a limiting case of 
a charged system state (\ref{Psim1e}). Indeed, when $q_2-q_1=0$,
wavefunction (\ref{Psim1e}) becomes extended (its norm tends to zero) 
and its coordinate dependence becomes identical to (\ref{coorK01}).  
Using Eq.~(\ref{Psim1-prob}) we deduce that the relative coordinate 
${\bf r} = {\bf r}_1 - {\bf r}_1 $ becomes maximally squeezed
(to the magnetic length $l_B=(\hbar c/qB)^{1/2}$ in the lowest LL)
at the expense of the center-of-charge coordinate 
${\bf R} = ( {\bf r}_1 + {\bf r}_1 )/2$. The latter becomes extended.
To elucidate this, let us consider {\em projections\/} of the 
center-of-charge $\bar{\bf R}$  and relative $\bar{\bf r}$  
coordinates onto zero LL.
In complex combinations of the components these are
$\bar{X} + i \bar{Y} = (B_e + B_h^{\dag})/\sqrt{2}$
and 
$\bar{x} + i \bar{y} = \sqrt{2}(B_e - B_h^{\dag}) = i \hat{K}_+$, respectively;  
here $\hat{K}_+ = \hat{K}_x + i \hat{K}_y$.
We see that (i) the projected relative and center-of-charge coordinates
become canonically conjugate, 
$ [\bar{x},\bar{Y}]=-[\bar{y},\bar{X}]=i$,
and (ii) the projected relative coordinate, up to a scaling factor and rotation by ninety degrees,
coincides with the MT operator $\hat{\bf K}$.

Let us discuss the relevant two-mode realization of the $SU(1,1)$ generators
\begin{eqnarray}
          \label{genK0}
     {\cal K}_0 &=&
   \frac{1}{2} \left( B_e^{\dag}B_e + B_h^{\dag}B_h  +  1 \right) \quad , \\
  {\cal K}_- &=&  B_e B_h   \quad , \quad
  {\cal K}_+ =  B_e^{\dag} B_h^{\dag}   \quad ,
\end{eqnarray}
which satisfy the $SU(1,1)$ commutation relations
$[ {\cal K}_{0} , {\cal K}_{\pm} ]  =  \pm {\cal K}_{\pm}$
and
$[ {\cal K}_{-}  , {\cal K}_{+}  ]    =   2 {\cal K}_{0}$.\cite{Perelomov86}
The dimensionless MT operator for a neutral magnetoexciton,
$\hat{\bf k}^2 = \hat{\bf K}^2 l_B^2/\hbar^{2}$, 
becomes in this representation
$\hat{\bf k}^2 =  2(2 {\cal K}_{0} - {\cal K}_{+} -  {\cal K}_{-})$,
while the angular momentum projection  
$\hat{L}_z =  B_h^{\dag} B_h  - B_e^{\dag} B_e$
is connected with the Casimir operator 
of the $SU(1,1)$ group\cite{Perelomov86} as
$\hat{C} =  ( \hat{L}_z^2 -1 )/4$.
This allows one to identify states (\ref{K0}) and (\ref{K})
as generalized coherent states of the $SU(1,1)$ group.\cite{Perelomov86}
Also, it becomes possible to find a representation of the MX
states in the set of ${\bf K}^2$ and $M_z$ quantum numbers, the eigenvalues
of the mutually commuting integrals of the motion $\hat{\bf K}^2$ and $\hat{L}_z$.
The corresponding state for, e.g., $M_z = - M \leq 0 $, is given by
\begin{equation}
        \label{KM}
  | {\bf K}^2,  M_z \rangle =
   e^{B_e^{\dag} B_h^{\dag}} | K ; M \rangle  \quad , 
\end{equation}
where 
\begin{equation}
        \label{KMcoh}
 | K; M \rangle  = \frac{1}{ \sqrt{I_M(k^2)} } \,
 \sum_{m=0}^{\infty}
\frac{ \left( \frac{ik_-}{\sqrt{2}} \right)^{m+M}
       \left( \frac{ik_+}{\sqrt{2}} \right)^{m}  }{ \sqrt{(m+M)!m!} } \,\,
          | m+M, m \rangle  
\end{equation}
is the Barut-Girardello coherent state,\cite{Barut71}  
$I_M(x)$ being a modified Bessel function
and 
$k_{\pm} = (K_x \pm i K_y) l_B/\hbar$.
Proof follows from Eqs.~(\ref{genK0}) and the algebra 
\begin{equation}
{\cal K}_{-} e^{- {\cal K}_{+} }  = e^{- {\cal K}_{+} }  
\left( {\cal K}_{+} +  {\cal K}_{-}  - 2 {\cal K}_{0} \right) \quad .
\end{equation}

In conclusion, we introduced an operator formalism  
for partial separation of degrees of freedom 
for electron-hole complexes in magnetic fields. 
We also established its connection with the $SU(1,1)$ algebras. 
Application of the powerful formalism of the $SU(1,1)$ group
allows one to construct a number of coherent ans squeezed 
states that may be useful for considering Coulomb correlations and 
optics of few- and many-particle systems in Landau levels.


\vspace*{6pt}


This work is supported in part by NSF grants DMR-0203560 and DMR-0224225, 
and by an award of Cottrell Research Corporation.



\end{document}